 \documentclass[aps,amsmath,amssymbd,dvips,nofootinbib,showkeys,preprint]{revtex4}

\usepackage{graphicx}
\usepackage{color}
\def \be  {\begin{equation}}
\def \ee  {\end{equation}}
\def \ee  {\end{equation}}
\def \bea {\begin{eqnarray}}
\def \eea {\end{eqnarray}}

\def \ra  {\rightarrow}

\begin{document}

\preprint{ECTP-2013-22\hspace*{0.5cm}and\hspace*{0.5cm}WLCAPP-2013-19}

\title{Degree of Chemical Non-equilibrium in Central Au-Au Collisions at RHIC energies}
\author{Abdel~Nasser~TAWFIK\footnote{http://atawfik.net/}}
%\email{a.tawfik@eng.mti.edu.eg}
\affiliation{Egyptian Center for Theoretical Physics (ECTP), Modern University for Technology and Information (MTI), 11571 Cairo, Egypt}
\affiliation{World Laboratory for Cosmology And Particle Physics (WLCAPP), Cairo, Egypt}

\author{ M.~Y.~El-Bakry}
\affiliation{Ain Shams University, Faculty of Education, Department of Physics, Roxi, Cairo, Egypt}

\author{ D.~M.~Habashy}
\affiliation{Ain Shams University, Faculty of Education, Department of Physics, Roxi, Cairo, Egypt}

\author{ M.~T.~Mohamed}
\affiliation{Ain Shams University, Faculty of Education, Department of Physics, Roxi, Cairo, Egypt}

\author{Ehab~ABBAS}
\affiliation{Egyptian Center for Theoretical Physics (ECTP), Modern University for Technology and Information (MTI), 11571 Cairo, Egypt}
\affiliation{World Laboratory for Cosmology And Particle Physics (WLCAPP), Cairo, Egypt}

\date{\today}

\begin{abstract}
    We investigate the difference between hadron resonance gas (HRG) calculations for chemical freeze-out parameters at fully and partly chemical equilibria. To this end, the results are compared with the particle ratios measured in central Au-Au collisions at a wide range of nucleon-nucleon center-of-mass energies, \hbox{$\sqrt{s_{NN}}=7.7-200~$GeV} as offered by the STAR experiment. We restrict the discussion to STAR, because of large statistics and overall homogeneity of STAR measurements (one detector) against previous experiments. We find that the matter produced at these energies is likely in fully chemical equilibrium, which is consistent with recent lattice QCD results. The possible improvements by partial chemical equilibrium ($\gamma_S\neq 1$) are very limited. We also discuss these results with the ones deduced from $\phi/\pi^-$ and $\Omega^-/\pi^-$ ratios. These hadron ratios are sensitive to the degree of chemical equilibrium. Accordingly, the conclusion that the matter produced reaches fully chemical equilibrium in central Au-Au at RHIC energies is confirmed. 
\end{abstract}


\maketitle

%%%%%%%%%%%%%%%%%%%%%%%%%%%%%%%%%%%%%%%%%%%%%%%%%%%%%%%%%%%%%%%%%%%%%%
%%%   Section I
%%%%%%%%%%%%%%%%%%%%%%%%%%%%%%%%%%%%%%%%%%%%%%%%%%%%%%%%%%%%%%%%%%%%%%

\section{Introduction}

Studying hadronic matter under extreme conditions of high temperature or density (or both) is the main proposal of experiments like ALICE at the Large Hadron Collider (LHC) and STAR at the Relativistic Heavy-Ion Collider (RHIC). The implementation of thermodynamic quantities in explaining the hadronization process in high-energy collisions dates back to more than six decades~\cite{Tawfik_Rev2014,Fermi,Hagedorn}. With the statistical bootstrap model, Hagedorn invented the concept of limiting temperature $T_{lim}$~\cite{Hagedorn}. Beyond this value, the hadronic matter is no longer stable. Accordingly, a new question arises, what will happen if $T \ge T_{lim}$? This was answered by Cabibbo and Parisi ~\cite{Cabibbo}. A new state of matter should be formed, the so called Quark-Gluon Plasma (QGP). Indeed, the Hagedorn temperature refers to the critical temperature related to the QGP-hadron phase transition. The heavy-ion collisions are conjectured as essential tools for creating QGP in laboratory. Such created hot and dense partonic matter rapidly expands and cools down. On path of this evolution, it undergoes phase transition(s) back to the hadronic matter. The particle ratios and transverse mass spectra  are important phenomenological indications for the thermal origin of the hadronization at high energies. The particle ratios and/or yields are studied in the framework of thermal models which success to explain it at various energies. 
  
There are many versions of these thermal models which can be classified in different ways. Depending on the degree of chemical non-equilibrium and/or the free parameters, the thermal models can be classified  into {\it fully} and {\it partially} chemical equilibrium and {\it non-equilibrium}.  
\begin{itemize}
\item In {\it fully} chemical equilibrium ~\cite{TAWFIK,Andronic,Cleymans,Tiwari,EVC}, only  two parameters are used to fit the experimental particle ratios. They are  $T_{ch}$ and $\mu_b$ being the freeze-out temperature and the chemical potential, respectively. 

\item In {\it partially} chemical equilibrium~\cite{Rafelski,Becattini,HRGpp,HRGpp2,HRGee}, an additional parameter is assumed, $\gamma_S$~\cite{Rafelski}. This parameter represents the degree of correctness of the assumption of the {\it absolute} chemical equilibrium. In other words, it  measures whether strangeness production was saturated in the full available phase space. The strange quark phase space occupation factor, $\gamma_S$, was needed in elementary collisions like proton-proton (pp)~\cite{HRGpp,HRGpp2} and electron-positron ($e^-e^+$)~\cite{HRGee}.  

\item In chemical {\it non-equilibrium}~\cite{Letessier}, two additional parameters are assumed. They are $\gamma_s$ and $\gamma_q$ being strange quark and light quark phase space occupation factor, respectively. Discussion on the possible physical meanings of these quantities can be found in Ref.~\cite{Gamma;means}.
\end{itemize}
  
In the present work, the three freeze-out parameters, $T_{ch}$, $\mu_b$ and $\gamma_S$, are extracted from fits of the experimental particle ratios with the corresponding ratios calculated in the HRG model, i.e., assuming partially chemical equilibrium. The experimental hadron ratios are limited to mid-rapidity central Au-Au collisions at energies ranging from $\sqrt{s_{NN}}~$ $200$ to $7.7~$GeV. This wide range of energies offers a good opportunity to check the degree of non-equilibrium in the most central Au-Au collisions measured by STAR experiment at RHIC energies. The reasons why we concentrate the study to the STAR  have been elaborated in Ref.~\cite{TAWFIK}. 

The results deduced in {\it partially} chemical equilibrium are compared with our previous study ~\cite{TAWFIK}, in which we assumed {\it fully} chemical equilibrium, i.e. $\gamma_S=1$. Also, they are confronted to recent lattice quantum chromodynamics (LQCD) predictions \cite{LQCD}. Furthermore, we shall discuss $\phi/\pi^-$ and $\Omega^-/\pi^-$ ratios. These two particle ratios are especially sensitive to degree of non-equilibrium in strange particle production. Finding out answers to the questions about the chemical equilibrium in this range of energy and the differences between the extracted parameters in partially and fully chemical equilibrium is the main goal of the present work, which is organized as follows. Section \ref{sec:hrg} elaborates details about the HRG model. The fits of the experimental ratios with the non-equilibrium HRG calculations are discussed in section \ref{sec:phys}. Section \ref{sec:res} is devoted to the results and discussion. The conclusions and outlook shall be summarized in section \ref{cons}.

\section{The hadron resonance gas model}
\label{sec:hrg}
     The HRG model was described by many authors~\cite{Becattini,HRG2003}. Here, we present a short summary with a special emphasis on partial chemical equilibrium. We starts with a partition function of a {\it free} gas consisting of hadron resonances. The partition function can be used to obtain the thermodynamic quantities. In hadronic phase and assuming fully chemical equilibrium ~\cite{Karsch:2003vd,Karsch:2003zq,Redlich:2004gp,Tawfik:2004sw,Tawfik:2004vv}
\bea \label{eq:lnz1}
\ln Z(T, \mu ,V)&=&\sum_i \ln Z^1_i(T,\mu_i ,V)=\sum_i\pm \frac{V g_i}{2\pi^2}\int_0^{\infty} p^2 dp \ln\left(1\pm \exp\left[\frac{\mu_i -\varepsilon_i(p)}{T}\right]\right),
\eea
where $\varepsilon_i(p)=(p^2+ m_i^2)^{1/2}$ is the $i-$th particle dispersion relation, $g_i$ is spin-isospin degeneracy factor and $\pm$ stands for bosons and fermions, respectively. The $i$-th particle chemical potential is given as $\mu_i=\mu_b B_i+\mu_s S_i+\mu_{I^3} I_i^3$, where $\ B_i$, $S_i$  and  $I_i^3$ are the baryon, strange and isospin quantum number, respectively. 
 
If we assume degree of the non-equilibrium, this can be implemented - among others - through the strange quark occupation factor $\gamma_S$ 
\bea \label{eq:lnzallG}
\ln Z(T, \mu, V,\gamma_S) &=& \sum_i\pm \frac{V\, g_i}{2\, \pi^2}\int_0^{\infty} p^2 dp\, \ln\left(1\pm \gamma_S^{s_i}\, \exp\left[\frac{\mu_i -\varepsilon_i(p)}{T}\right]\right),
\eea 
where $s_i$ is the number of strange valence quarks and antiquarks in the $i$-th hadron. 
At finite temperature $T$ and chemical potential $\mu_i$, the pressure of the $i$-th hadron or resonance species reads 
\begin{equation}
\label{eq:prss} 
p(T,\mu_i ,\gamma_S) = \pm \frac{g_i}{2\pi^2}T \int_{0}^{\infty}
           p^2 dp  \ln\left(1 \pm \gamma_S^{s_i} \exp\left[\frac{\mu_i -\varepsilon_i(p)}{T}\right]\right).
\end{equation} 
As no phase transition is conjectured in HRG, summing over all hadron resonances results in the final thermodynamic pressure. Accordingly, the number density can be obtained as
\bea \label{eq:n_i}
n(T, \mu,\gamma_S) &=& \sum_i \frac{\partial}{\partial \mu_i} p(T,\mu_i,\gamma_S).
\eea 

The conservations laws should be fulfilled through the chemical potentials and temperature and over the whole phase space. So, $\mu_s$ and $\mu_{I^3}$ can be calculated at a certain $T$ and $\mu_b$.

During the expansion of the hadronic matter, we assume that inelastic interactions between resonances and annihilation processes~\cite{SHMUrQM} have negligible contributions in the final state. The main process at this stage is unstable resonance decay. With this regard we recall the so-called {\it ''proton anomaly''} \cite{prtnanom}. One of possible scenarios is based on some of these two mechanisms. Nevertheless, the anomaly still there unsolved, in a solid way. The main process at this stage is unstable resonance decay. So, the final number density reads
\bea \label{eq:n_i^{final}}
\ n_i^{final} &=& n_i + \sum_j\ Br_{j\ra i}\; n_j,
\eea 
where $Br_{j\ra i}$ is the effective branching ratio of $j$-th hadron resonance into $i$-th particle. Taking into consideration all multi-step decay cascades, then 
\bea \label{eq:Br}
\ Br_{j\ra i} &=& br_{j\ra i} + \sum_{l_1}\ br_{j\ra l_1}br_{l_1\ra i}+ \sum_{l_1,l_2}\ br_{j\ra l_1}br_{l_1\ra l_2} br_{l_2\ra i}+ \cdots,
\eea 
where the $br_{j\ra i}$ is the number of $i$-th particle which is produced from the decay of $j$-th particle.
    
In the present work, we include contributions of the hadrons which consist of light and strange quark flavors, only \cite{Karsch:2003vd,Karsch:2003zq,Redlich:2004gp,Tawfik:2004sw,Tawfik:2004vv}. These are listed in the most recent PDG~\cite{PDG}. The branching ratios are also taken from Ref.~\cite{PDG} with zero-width approximation. The Excluded-Volume Correction (EVC)~\cite{EVC} is applied taking into account the volume occupied by individual hadrons with radii $r_m$ for mesons and $r_b$ for baryons. The thermodynamic quantities are conjectured to be modified due to EVC. Accordingly, the corrected pressure will be obtained by an iterative procedure, 
 \bea 
 p^{excl}(T, \mu_i) &= p^{id}(T, \tilde{\mu_i}), ~~~~~~~~~~~{\tilde{\mu_i}} &= \mu_i - \upsilon p^{excl}(T, \tilde{\mu_i}), \label{eq:P}
\eea 
where $p^{id} (p^{excl})$ being the pressure in the ideal case (case of excluded volume) and $\upsilon$  is the eigen volume which is calculated for a radius, $16 \pi r^3/3$~\cite{Landau}.  We would like to stress that the details about hadron resonances mentioned above as well as the experimental ratio sets (given in section \ref{sec:phys}) are the same as the ones used in our previous analysis~\cite{TAWFIK}. We keep these unchanged in order to compare the chemical equilibrium with the present case.

%%%%%%%%%%%%%%%%%%%%%%%%%%

\section{Statistical Fits with STAR particle ratios}
\label{sec:phys}   
    
The criterion for the best statistical fit is based on assuring minima, for instance
\bea \label{eq:chi}
\chi^2 &=& \sum_i\frac{\left(R_i^{exp}-R_i^{model}\right)^2}{\sigma^2},
\eea 
where $R_i^{exp}$ ($R_i^{model}$) being the $i$-th measured (calculated) ratio and $\sigma$ is the experimental data errors.     

\begin{description}
\item{\bf At $200~$GeV}, we use yields of pions, kaons, (anti)protons~\cite{STAR2}, $\Lambda$, $\bar{\Lambda}$ and multi-strange baryons~\cite{STAR3,STAR4} measured at mid-rapidity measured in  the STAR experiment at centrality $0-5\%$, except the $\Omega$/$\bar{\Omega}$ ratios are measured at $0-20\%$~\cite{STAR4}. The measured pions spectra are corrected for feed-down from weak decays as well as $\Lambda$ ($\bar{\Lambda}$) are corrected for feed-down from weak decays of multi-strange baryons.  

\item{\bf At $130~$GeV and $62.4~$GeV}, we use yields of pions, kaons, (anti)protons~\cite{STAR2}, $\Lambda$, $\bar{\Lambda}$ and multi-strange baryons~\cite{STAR5,STAR4,STAR6} measured at mid-rapidity measured in the STAR experiment at centrality $0-5\%$ except the multi-strange baryons have been measured at $0-20\%$. The measured pions spectra are corrected for feed-down from weak decays. $\Lambda$ ($\bar{\Lambda}$) are corrected for feed-down from weak decays of $\Xi$ at $62.4~$GeV. 

\item{\bf At $39$, $11.5$ and $7.7$ GeV}, we use for yields of pions, kaons, (anti)protons~\cite{STAR7}, $\Lambda$, $\bar{\Lambda}$,  $\Xi$ and $\bar{\Xi}$~\cite{STAR8} measured at mid-rapidity measured in the STAR experiment at centrality $0-5\%$. The $\Omega$/$\bar{\Omega}$ ratios at centrality $0-5\%$, $0-20\%$ and $0-60\%$ for $39$, $11.5$ and $7.7$ GeV, respectively, are taken from Ref.~\cite{privateCommunication}. The measured pion spectra have been corrected for feed-down from weak decays as well as $\Lambda$ ($\bar{\Lambda}$) for the feed-down contributions from $\Xi$ weak decay. The analysis includes the available $10$ independent ratios where $\Omega/\pi$ is not included. 
 
\end{description}

At $200$, $130$ and $62.4~$GeV, the set of particle ratios used in the present analysis  are $\pi^-$/$\pi^+$, $k^-$/$k^+$, $\bar{p}/p$, $\bar{\Lambda}$/$\Lambda$, $\bar{\Xi}$/$\Xi$, $\bar{\Omega}$/$\Omega$, $k^-$/$\pi^-$, $\bar{p}$/$\pi^-$, $\Lambda$/$\pi^-$, $\Xi$/$\pi^-$ and $\Omega$/$\pi^-$ except for $200$ GeV, we also use $(\Omega+\bar{\Omega})/\pi^-$ instead of $\Omega$/$\pi^-$.  

At $39$, $11.5$ and $7.7$ GeV, the particle ratios $\pi^-$/$\pi^+$, $k^-$/$k^+$, $\bar{p}$/p, $\bar{\Lambda}$/$\Lambda$, $\bar{\Xi}$/$\Xi$, $\bar{\Omega}$/$\Omega$, $k^-$/$\pi^-$, $\bar{p}$/$\pi^-$, $\Lambda$/$\pi^-$ and $\bar{\Xi}$/$\pi^-$ are used\footnote{The experimental ratios at $39$, $11.5$ and 7.7 GeV, $k^+$/$\pi^+$ and p/$\pi^+$, are calculated using the other experimental ratios assuming the same relative error in $k^-$/$\pi^-$ and $\bar{p}$/$\pi^-$, respectively.}.

\begin{figure}[htb]
\centering{
\includegraphics[width=7.5cm]{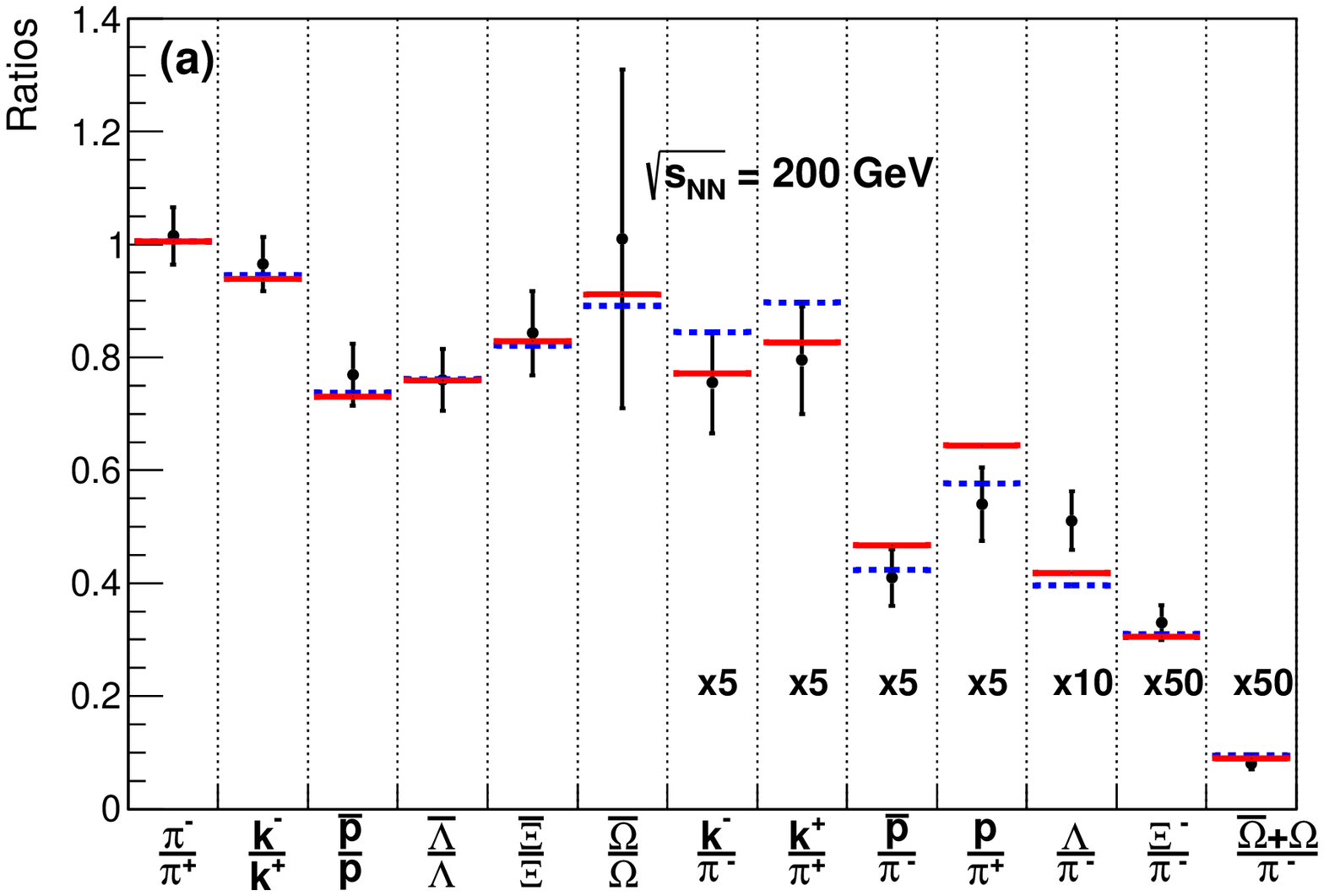}
\includegraphics[width=7.5cm]{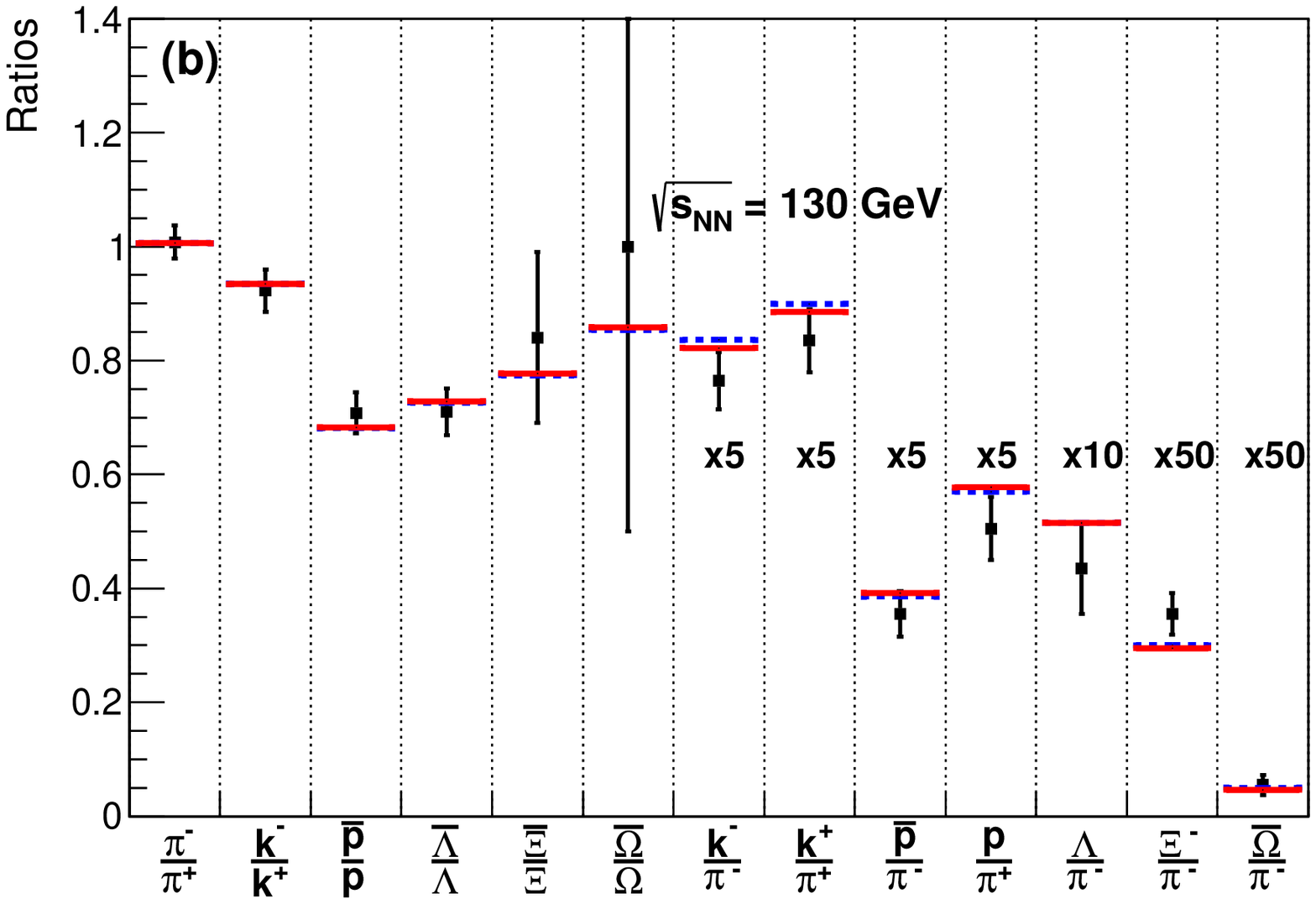}
\caption{(Color online) Left-hand panel (a) The experimental particle ratios (symbols) ~\cite{STAR2,STAR3,STAR4} are compared to the HRG calculations assuming partially chemical equilibrium (horizontal solid lines) and assuming fully chemical equilibrium (horizontal dash lines) at 200 GeV  where HRG calculations was preformed at $T_{ch}$ and $\mu_b$ parameters which assure minimum $\chi^2$ per degrees-of-freedom. For comparison with other ratios and better appearance, some ratios are scaled (scaling factor are given) so that we avoid log plotting. Right-hand panel (b) shows the same as in (a) but at $130~$GeV, where the experimental particle ratios are taken from Refs.~\cite{STAR2,STAR5,STAR4,STAR6} 
 \label{fig200_130} }
}
\end{figure}

\begin{figure}[htb]
\centering{
\includegraphics[width=7.5cm]{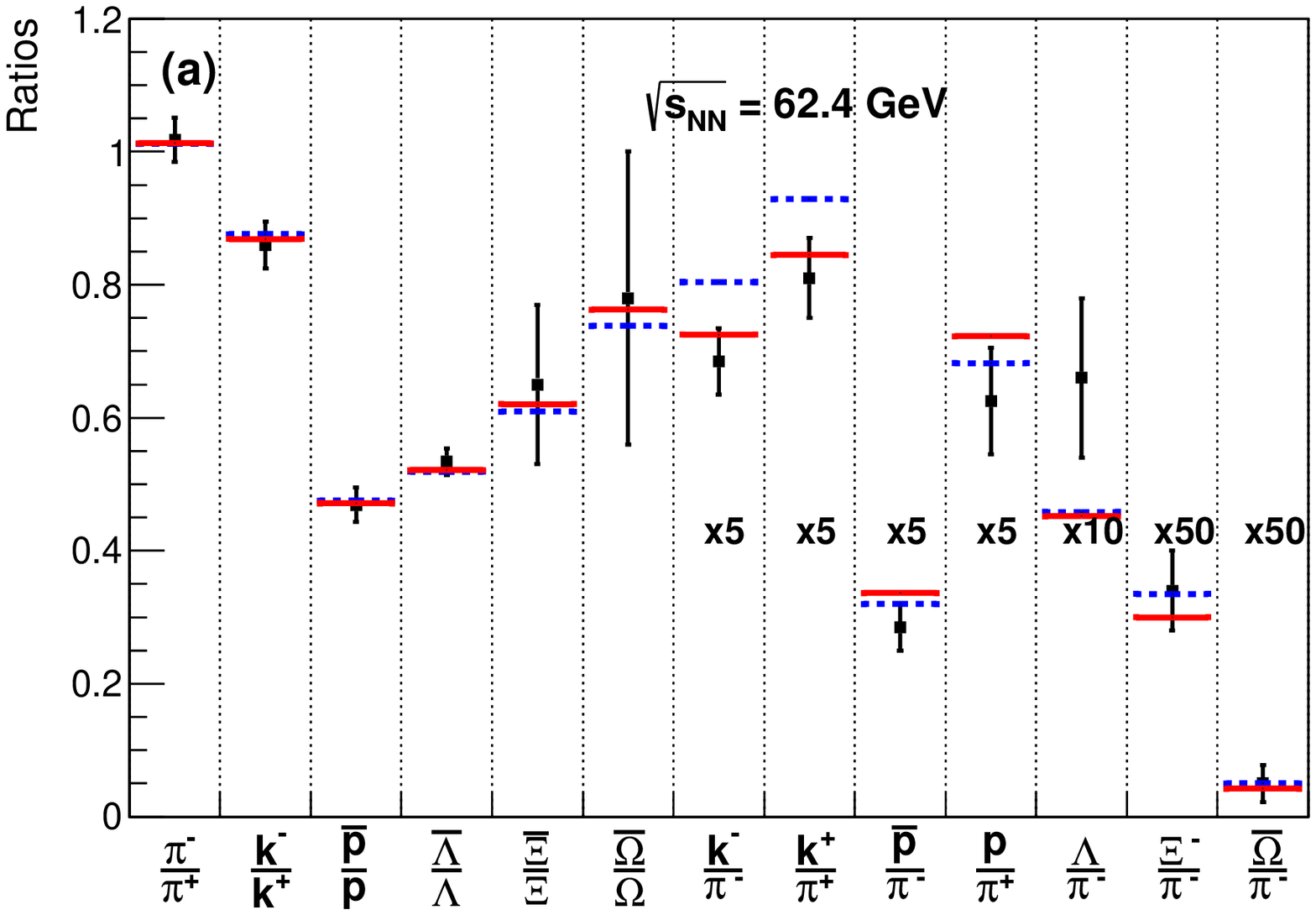}
\includegraphics[width=7.5cm]{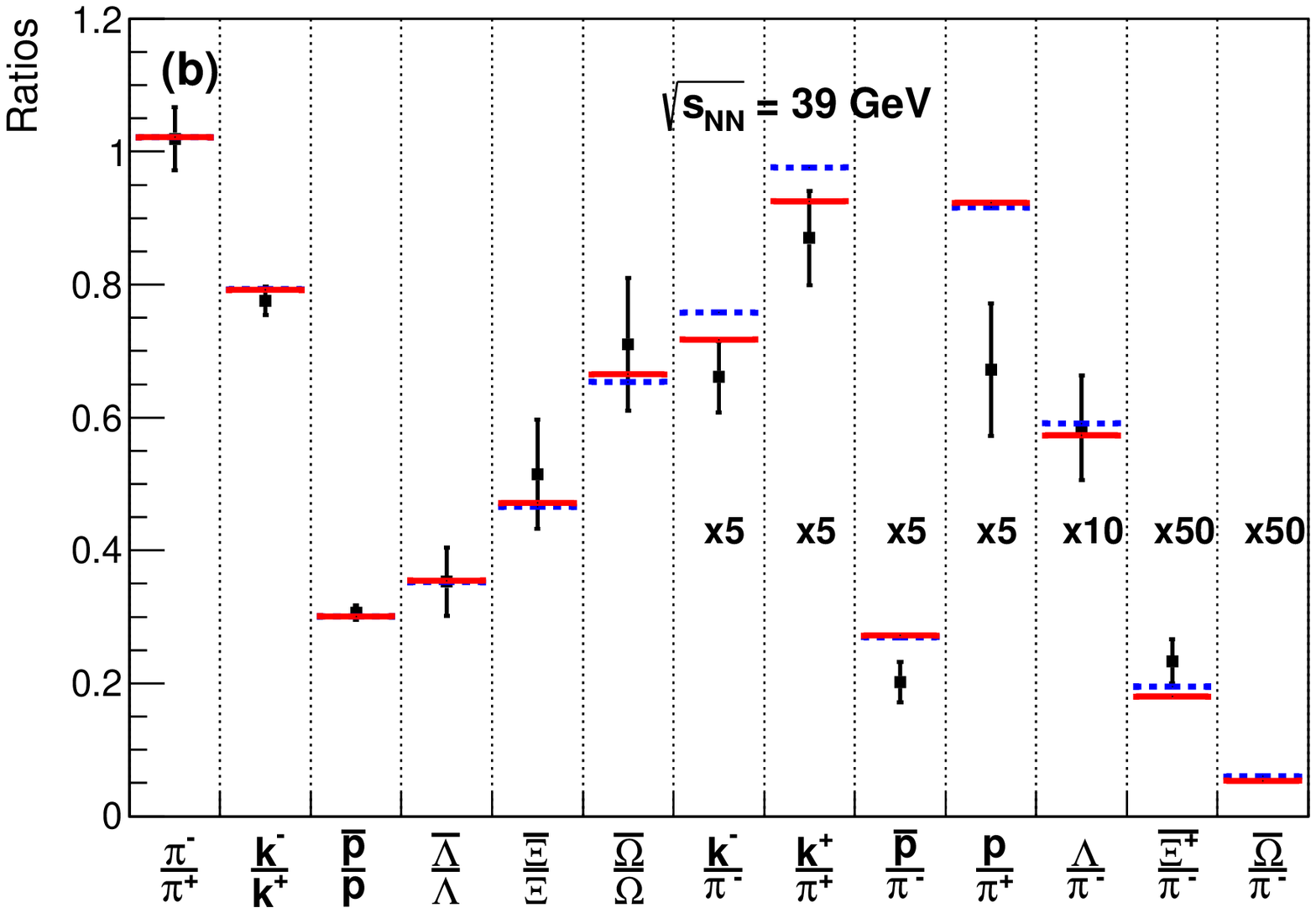} \\
\includegraphics[width=7.5cm]{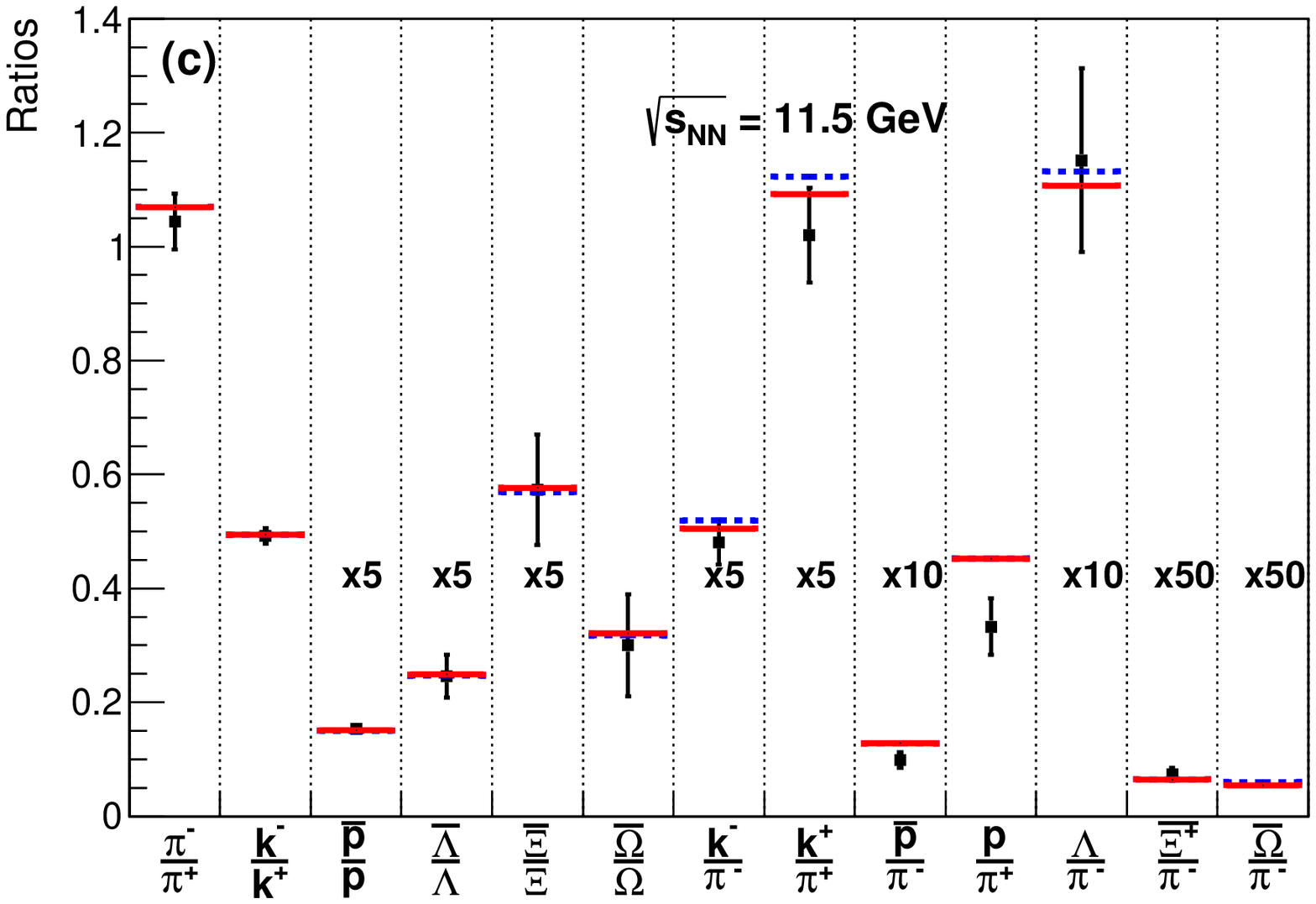}
\includegraphics[width=7.5cm]{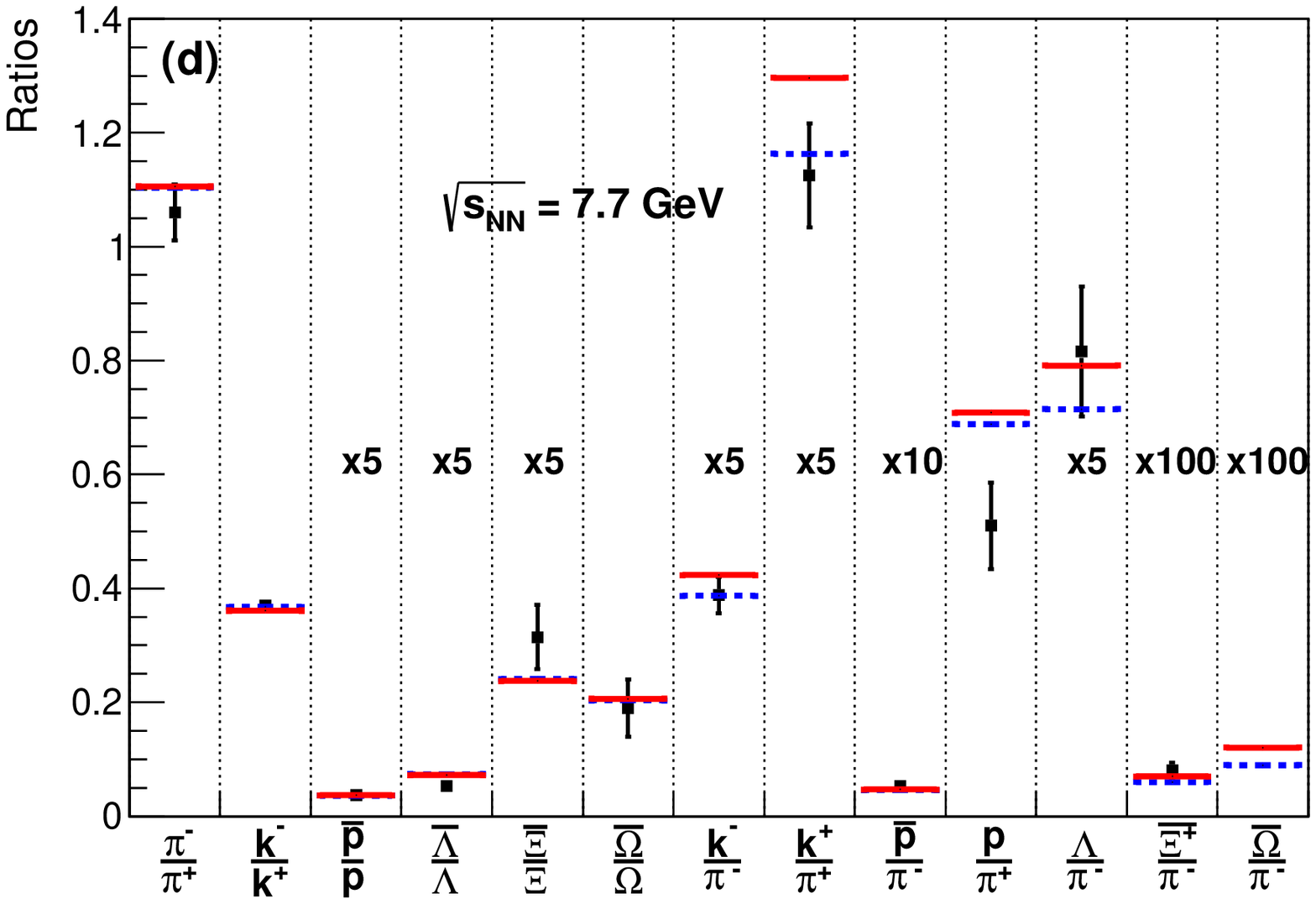} 
\caption{(Color online) The same as in Fig. \ref{fig200_130} but at lower STAR-BES energies (see text for details). 
 \label{figBES} 
}}
\end{figure}

The comparison between the experimental (symbols), calculated particle ratios at $\gamma_S \neq 1$ (horizontal lines) and the ones at $\gamma_S=1$~\cite{TAWFIK} (horizontal dashed lines) is given in Figs. \ref{fig200_130} and \ref{figBES} at various energies. We note that the partially chemical equilibrium causes a little improvement in reproducing the various particle ratios, especially the strange ones. Further results and highlights shall be elaborated in section \ref{sec:res}.

\section{Results and Discussion}
\label{sec:res}

\begin{table}[htb]
%\centering
\begin{tabular}{|c|  | c    | c    | c    || c   | c    | c | c |}     
 % centered columns (6 columns)
\hline
 & \multicolumn{3}{|c||}{$\gamma_S =1$~\cite{TAWFIK}} &\multicolumn{4}{|c|}{$\gamma_S \neq 1$~[present work]}\\
\hline     %inserts double horizontal lines
 $\sqrt{s_{NN}}$ [GeV] & $T_{ch}$  & $\mu_b$ & $\chi^2/dof$ & $T_{ch}$  & $\mu_b$ & $\gamma_S$ & $\chi^2/dof$ \\ [0.5ex] % inserts table      %heading
\hline       % inserts single horizontal line
200 & 159.5 & 27.5 & 9.244/9 & 167 & 30 & 0.9 & 7.101/8  \\   % inserting body of the table
130 & 157  & 34 & 7.072/9 & 158  & 34 & 0.98 & 6.95/8  \\
62.4 & 157 & 66 & 10.512/9 & 161.5 & 68.5 & 0.89 & 6.91/8 \\
39 & 160.5 & 110.5 & 11.03/8 & 162 & 111 & 0.94 & 10.186/7  \\
11.5 & 153 & 308 & 6.283/8  & 153.5 & 308 & 0.97 & 6.159/7 \\ 
7.7 & 145 & 409 & 15.074/8 & 144.5 & 413 & 1.12 & 12.148/7    \\ [1ex]
 % [1ex] adds vertical space
\hline
 %inserts single line
\end{tabular}
\caption{The freeze-out parameters, $T_{ch}$, $\mu_b$ and $\gamma_S$ are estimated from $\chi^2$ fitting approaches assuming chemical equilibrium $\gamma_S =1$~\cite{TAWFIK} and partially chemical equilibrium $\gamma_S \neq 1$ using single hard-core for mesonic and baryonic ($r_m=r_b=0.3~$fm) constituents of the HRG model. The experimental particle ratios are measured in STAR experiment at various energies, i.e., first phase of the beam energy scan (BES) program.
\label{tab:1}
}
\end{table}

%\subsection{Comparing fully with partially chemical-equilibrium} 

The freeze-out parameters, $T_{ch}$, $\mu_b$ and $\gamma_S$ are estimated from $\chi^2$  fitting approaches assuming finite hard-core (single hard-core radius, $r_m=r_b=0.3~$fm) constituents of the HRG model. They are listed out in Tab. \ref{tab:1}. 
Some remarks are now in order.
\begin{itemize}
\item We observe that $\chi^2/dof$ values at fully chemical equilibrium, i.e., $\gamma_S =1$ are less than the ones at partially chemical equilibrium, i.e. $\gamma_S \neq 1$, at $130$, $39$ and $11.5~$GeV. 
\item In fact, by making a scan of a three-parameters fit at $200~$GeV, the best fit occurs at  $\gamma_S=0.9$, with $\chi^2/dof$= 7.101/8, i.e., only $0.139$ less than the fit for $\gamma_S =1.0$. On the other hand, at $62.4~$GeV, we obtain a better fit when $\chi^2/dof = 6.91/8$  i.e., $0.30$ less than the fit at $\gamma_S =1.0$. Slight improvements cannot be seriously taken as a proof that the partially equilibrium is more suitable. 
\item The freeze-out parameters deduced from the lattice QCD calculations show that the temperature is nearly constant at high energies, i.e. small chemical potentials~\cite{LQCD}. On the other hand, we found that the parameters $T_{ch}$ and $\mu_b$ at fully chemical equilibrium\cite{TAWFIK} are more consistent with the lattice QCD than the others extracted from partially chemical equilibrium ($\gamma_S \neq 1$), Fig. \ref{fig:TmuB_comp}. Here, the freeze-out parameters at partially chemical equilibrium ($\gamma_S \neq 1$) are slightly larger than the ones at fully chemical equilibrium, especially temperature. This effect can be also seen~\cite{Andronic} at higher SPS energies. 
 
\begin{figure}[htb]
\centering{
\includegraphics[width=5.0cm,angle=-90]{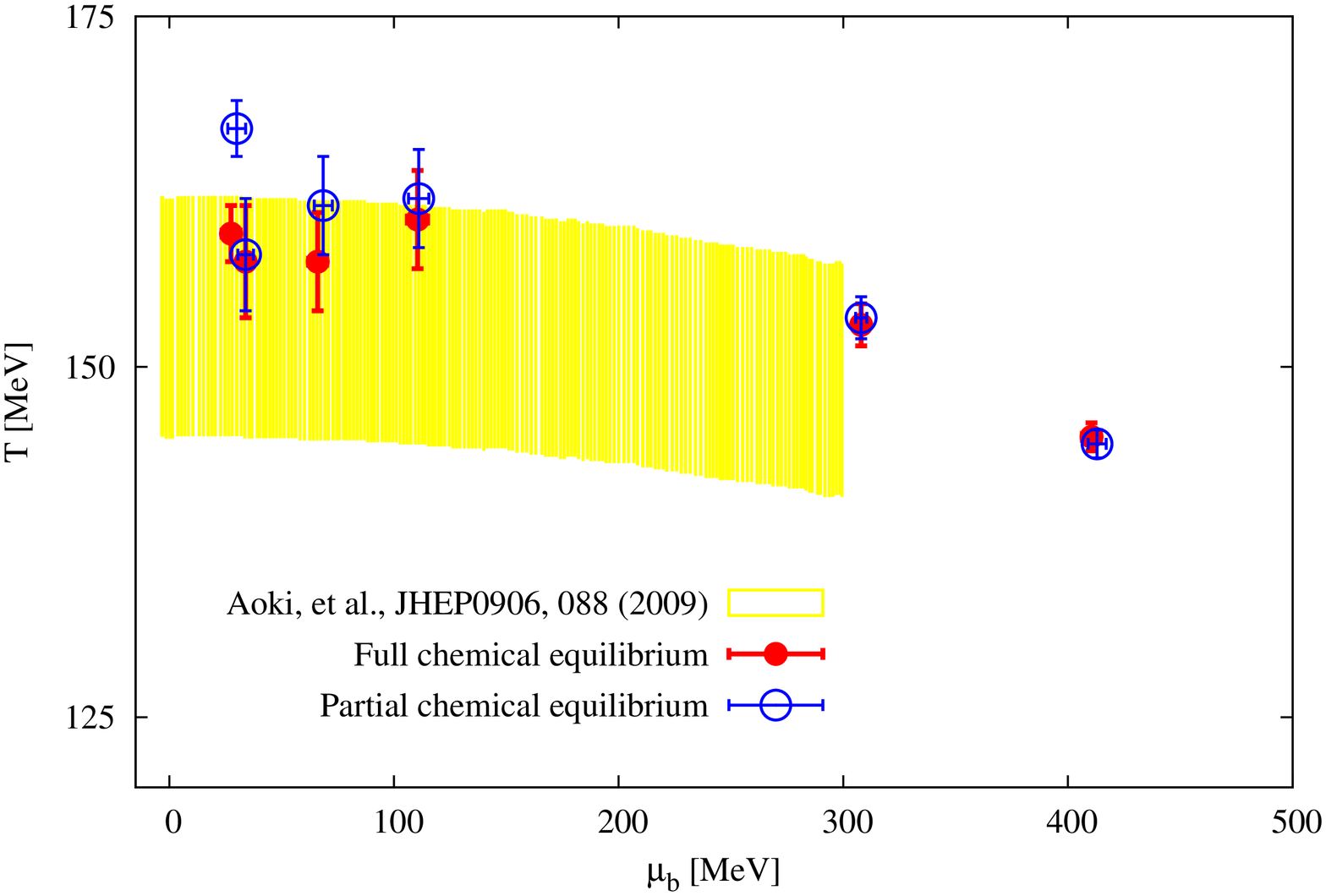}
\includegraphics[width=5.0cm,angle=-90]{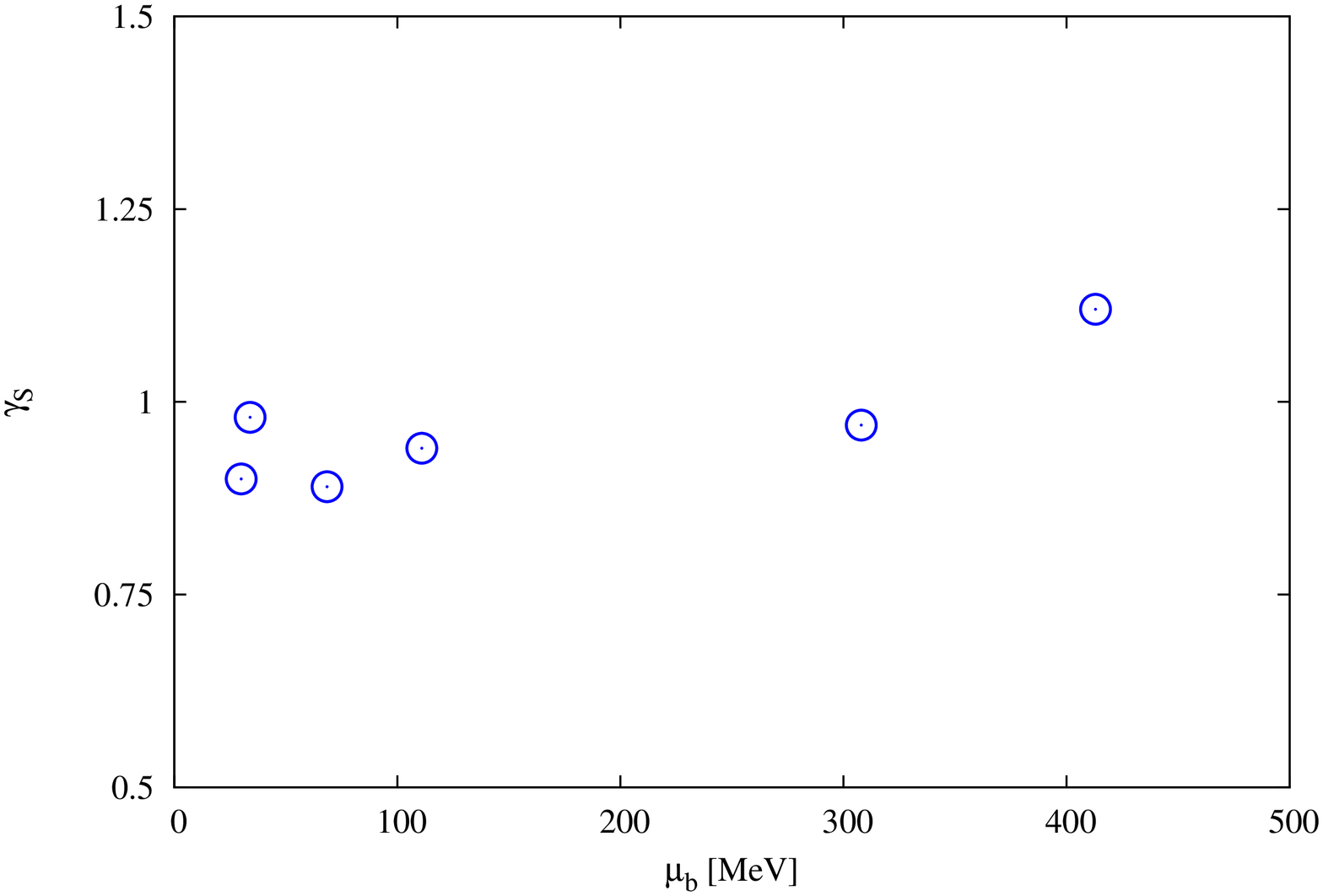}
\caption{(Color online) Left-hand panel: the freeze-out parameters assuming fully chemical equilibrium~\cite{TAWFIK} (empty symbols) and assuming partially chemical equilibrium (full symbols) are compared with lattice QCD calculations \cite{LQCD} (band). The right-hand panel shows the dependence of $\gamma_S$ on the chemical potential (related to center-of-mass energy), Tab. \ref{tab:1}.
 \label{fig:TmuB_comp} }
 }
\end{figure}

\item The energy-dependence of $\gamma_S$ seems to be non-monotonic, Tab. \ref{tab:1}. At $7.7~$GeV, $\gamma_S$ gets a value larger than unity, $1.12$. These results is contract with  that was predicted~\cite{Becattini2,Kampfer} that $\gamma_S$ remains smaller than unity and increases to unity as energy increases.  The problem of estimating $\gamma_S$ at low energy is very obvious. Fortunately, the present work is devoted to {\it ''high hnergy''}, analysis of the critical temperature, RHIC top energies. 
\item Finally, we notice that the parameters at partially chemical equilibrium are accompanied by some modifications in the particle ratios, especially in $K/\pi$. This might be originated in the advantages of taking into consideration the new parameter, $\gamma_S$, which apparently has a strong effect on the strange particles. 
\end{itemize}

\subsection{Freeze-out insights from $\Omega/\pi^-$ and $\phi/\pi^-$ ratios}

\begin{figure}[htb]
\centering{
\includegraphics[width=10.0cm,height=7.0cm,angle=0]{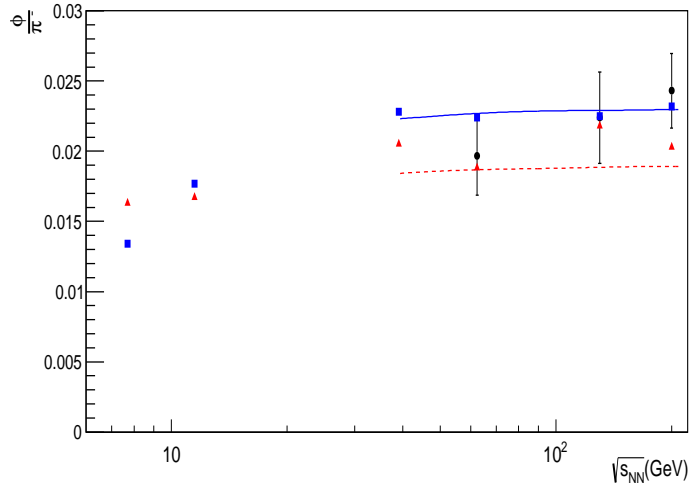}
\caption{(Color online) The experimental ratios of $\phi/\pi^-$ \cite{phi1,phi2} (circles) are compared with HRG calculation at fully (squares) and partially chemical equilibrium (triangles). The curves represent the HRG calculations at a fixed $T_{ch}=158.5~$MeV and $\gamma_S=1$ (solid curve) and  $\gamma_S=0.9$ (dashed curve). It should be noted that the experimental data are collected at different centralities (see text).
 \label{fig:Phipi}} 
}
\end{figure}

In framework of HRG, the pion number density and other thermodynamic quantities at $\sqrt{s_{NN}}>30~$ GeV was conjectured to be constant~\cite{Andronic09;12}. Thus, the change in the particle ratios $\Omega/\pi^-$ and $\phi/\pi^-$ is characterized by the changed number density of $\Omega$ and $\phi$, respectively. Starting with  $\phi/\pi^-$, the thermal number density of $\phi$ can be written as (in Boltzmann limit)
\bea \label{n}
n_{\phi}= \gamma_S^2\;\frac{T g_{\phi}}{2 \pi^2}\; m_{\phi}^2 \; K_2\left(\frac{m_{\phi}}{T}\right) ,
\label{denphi}
\eea 
where $g_{\phi}$ and $m_{\phi}$ are degeneracy and mass of $\phi$ meson, respectively. $\phi/\pi^-$ ratio is not included in extracting the freeze-out parameters in both cases.  From the above equation, $\phi$ meson is sensitive to both temperature and $\gamma_S$. Furthermore, $\phi/\pi^-$ was used to check whether the strangeness suppression in pp interactions according to the canonical suppression or according to $\gamma_S$ ~\cite{HRGpp}. The $\phi$-meson is not canonically suppressed, because it carries zero strangeness quantum number (strange and anti-strange quark). Therefore, this ratio can be used to determine {\it the degree of chemical non-equilibrium} in high-energy collisions. In Fig. \ref{fig:Phipi}, we confront the HRG calculations in fully and partially chemical equilibrium with the experimental data~\cite{phi1,phi2} of Au-Au collision at energies $200$ ,$130$ and $62.4~$ GeV at centrality $0-5\%$, $0-11\%$ and $0-20\%$, respectively. Although, this is apparently based on the observation that the HRG calculations at fully and partially chemical equilibrium are almost indistinguishable, the fitting with partially chemical equilibrium has a slightly (or even no) improvement in $\chi^2/dof$. Furthermore, this fitting results in non-monotonic energy-dependence of $\gamma_S$ as well as $T_{ch}$ , Fig. \ref{fig:TmuB_comp}.
 
When $\phi$-meson is not depend on $\mu_b$  and one use the approximately constant temperature at energies $\sqrt{s_{NN}}>39~$GeV \cite{TAWFIK}, a small variation in $\gamma_S$ will be clear in experimental data. To guide the eyes, we draw two curves representing the HRG calculations at fixed $T_{ch}=158.5~$MeV,as limiting temperature $T_{ch}=158.5~$MeV estimated in Ref. ~\cite{TAWFIK}, assuming $\gamma_S = 1$ (solid curve) and  $\gamma_S =0.9$ (dashed curve). The experimental data the $\phi/\pi^-$ seems to be approximately constant which lead that $\gamma_S$ should be constant or vary very slowly. 

It is worthwhile to highlight here that the two curves almost equally agree  with the experimental data. Nevertheless, the curve at fully chemical equilibrium assures all physical aspects discussed in previous paragraphs.

\begin{figure}[htb]
\centering{
\includegraphics[width=10.0cm,height=8.0cm,angle=0]{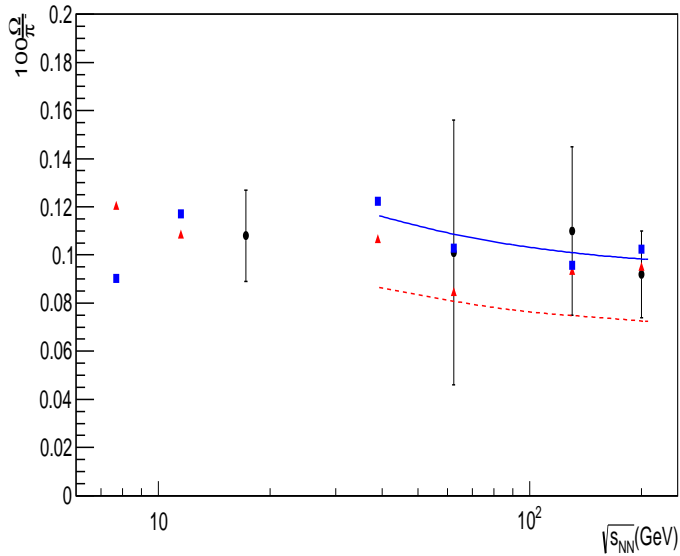}
\caption{(Color online) The experimental ratios of $\Omega/\pi^-$ (circles), the SPS results are from NA57~\cite{NA57} (centrality $0 - 11\%$) and the RHIC values are from STAR~\cite{STAR5,STAR4,STAR6} (centrality $0 - 20\%$), are compared with HRG assuming fully (squares) and partially chemical equilibrium (triangles). The two curves represent the HRG calculations at fixed $T_{ch} =158.5~$MeV assuming $\gamma_S=1$ (solid) and  $\gamma_S=0.9$ (dashed). 
 \label{fig:Omegapi}}
}
\end{figure}

In Boltzmann limit, the number density of $\Omega$ baryon can be determined as
\bea \label{n}
n_{\Omega}= \gamma_S^3{{Tg_{\Omega}}\over {2\pi^2}}\; m_{\Omega}^2\;  K_2\left({{m_{\Omega}}\over T}\right) \; \exp\left(\frac{\mu_b-3\mu_s}{T}\right),
\label{denomega}
\eea
where $g_{\Omega}$ and $m_{\Omega}$ are degeneracy and mass of $\Omega$ baryon, respectively.  It is obvious that $\Omega/\pi^-$ ratio is very sensitive to  $\gamma_S$ as well as to the temperature. The temperature-dependence of this ratio was discussed in Ref.~\cite{Andronic}. Fig. \ref{fig:Omegapi} shows that the experimental ratios~\cite{STAR5,STAR4,STAR6,NA57}  compared to calculations from the HRG model at fully and partially chemical equilibrium. As done in Fig. \ref{fig:Phipi}, two curves representing HRG calculations at fixed $T_{ch} =158.5~$MeV \cite{TAWFIK} and $\gamma_S = 1$ (solid curve) and $\gamma_S =0.9$ (dashed curve). The fit with at fully chemical equilibrium seems to indicate expandability of extra-parameters like $\gamma_S\neq 1$. 

Here, the experimental data of $\Omega/\pi^-$ seems to be constant. If one would assume that the temperature varies very slowly at relatively low chemical potential as predicted by LQCD~\cite{LQCD} and HRG calculations~\cite{TAWFIK}, one would be able to deduce from Eq. (\ref{denomega}) that $\gamma_S$ should remain fixed in this range of energy or allows it to vary but very slowly.

More accurate measurements for  $\Omega/\pi^-$ and $\phi/\pi^-$ in relative high energy (with the help of approximately constant temperature), can help to make a non-doubtful conclusion about the degree of chemical equilibrium at high energy.

\section{Conclusions}
\label{cons}

We have introduced a systematic analysis to perform an accurate discrimination of the quality of the fits with and without $\gamma_S$, using particle ratios instead of yields. Several authors pointed out that this method introduces potentially severe biases in the parameters and the $\chi^2$ and in fact results in an inaccurate assessment of the fit quality \cite{Manninen} 
Thus, we want to present argumentation in favor of the implementation of particle ratios in order to deduce the freezeout parameters. \begin{itemize}
\item In general ,the particle-antiparticle ratios have low errorbars relative to all other particle ratios and of course yields. This is because the effect of rapidity cut will be cancelled out, the same as for flow. Furthermore, the corrected baryon-antibaryon ratios are resonance independent.
\item Form statistical point-of-view, the present analysis finds on TWO variables, $T$ and $\mu$, by fixing $11$ different ratios. Obviously, this would more accurate than finding THREE variables by fixing even $12$ particle ratios.
\item The previously mentioned extra variable, $d V/d y$ in mid-rapidity, which was needed to deal with the yields instead of the particle ratio seems no to have an exact definition. 
\end{itemize}

Beside these argumentations, we may also highlight that the ratios likely reduce the volume fluctuations, while the quantity $d V/d y$ surely come up with additional contribution to uncertainty. Despite all these argumentations, we should not oversee any possible  introduction of potential severe biases to the freezeout parameters and the $\chi^2$.

We have investigated the differences between HRG calculations for the chemical freeze-out parameters at fully and partially chemical equilibria. To this end, the HRG results are compared with the particle ratios measured in central Au-Au collisions at a wide range of energies, \hbox{$\sqrt{s_{NN}}=7.7-200~$GeV} as offered by the STAR BES-I. We conclude that the matter produced at these energies is likely in fully chemical equilibrium. This is also consistent with recent lattice QCD results. The possibility to improve the results by partially chemical equilibrium ($\gamma_S\neq 1$) are very limited. Actually, there is no strong justification for the time-consuming calculations for $\gamma_S$ as an additional freeze-out parameter at least in the central collisions analysed in the present work.

   Although, this is apparently based on the observation that the HRG calculations at fully and partially chemical equilibrium are almost indistinguishable in the figures, the fitting with partially chemical equilibrium have slightly (or even no) improvement in $\chi^2/dof$. Furthermore, it leads to non-monotonic energy-dependence  in  $\gamma_S$ and temperatures 

 A large variation in $\Omega/\pi^-$ and $\phi/\pi^-$ would be seen in experimental data, if $\gamma_S$ is allowed to vary with the energy, especially at approximately fixed temperature. As seen from Figs. \ref{fig:Omegapi} and \ref{fig:Phipi}, when there is a slightly (or even no) improvement through the fitting according to the additional parameter $\gamma_S$, the fully chemical equilibrium becomes more acceptable. 
 
The results in the present work also consistent in general with with previous study in~\cite{Manninen,Kampfer,Kaneta} at $200$ and $130~$GeV at most central Au-Au collision. In these studies the  $\gamma_S$ reached unity or very close to it at the central Au-Au collisions. Accordingly, the outcome that the matter produced reaches full chemical equilibrium in central Au-Au at RHIC energies is found. 
 
Again, the EVC account for the volume occupied by individual hadrons with radii $r_m$ for mesons and $r_b$ for baryons. The resulting freeze-out parameters $T_{ch}$, $\mu_b$ and even $\gamma_S$ refer to irrelevance of EVC, especially for single-core radii $r_m=r_b \leq 0.3~$fm.
 
%%%%%%%%%%%%%%%%%%%%%%%%%%%%%%%%%%%%%%%%%%%%%%%%%%%%%%%%%%%%%%%%%%%%%%
%%%   References
%%%%%%%%%%%%%%%%%%%%%%%%%%%%%%%%%%%%%%%%%%%%%%%%%%%%%%%%%%%%%%%%%%%%%%

%-----------------------------------------------

%----------------------==================-------------------------

\end{document}